# Inverse Orbital Torque via Spin-Orbital Entangled States

E. Santos[1], J. E. Abrão[1], Dongwook Go[2,3], L. K. de Assis[4], Yuriy Mokrousov[2,3], J.B.S. Mendes[5], and A. Azevedo[1]

[1] *Departamento de Física, Universidade Federal de Pernambuco, Recife, Pernambuco 50670-901, Brasil.*
[2] *Peter Grünberg Institut and Institute for Advanced Simulation, Forschungszentrum Jülich and JARA, 52425 Jülich, Germany.*
[3] *Institute of Physics, Johannes Gutenberg University Mainz, 55099 Mainz, Germany.*
[4] *Universidade Federal de Pernambuco, Programa de Pós-Graduação em Ciências dos Materiais, Recife, Pernambuco 50740-560, Brasil.*
[5] *Departamento de Física, Universidade Federal de Viçosa, 36570-900 Viçosa, Minas Gerais, Brazil.*

While current-induced torque by orbital current has been experimentally found in various structures, evidence for its reciprocity has been missing so far. Here, we report experimental evidence of strong inverse orbital torque in YIG/Pt/CuO$_x$ (YIG = Y$_3$Fe$_5$O$_{12}$) mediated by spin-orbital entangled electronic states in Pt. By injecting spin current from YIG to Pt by the spin pumping via ferromagnetic resonance and by the spin Seebeck effect, we find a pronounced inverse spin Hall effect-like signal. While a part of the signal is explained as due to the inverse spin-orbital Hall effect in Pt, we also find substantial increase of the signal in YIG/Pt/CuO$_x$ structures compared to the signal in YIG/Pt. We attribute this to the inverse orbital Rashba-Edelstein effect at Pt/CuO$_x$ interface mediated by the spin-orbital entangled states in Pt. Our work paves the way toward understanding of spin-orbital entangled physics in nonequilibrium and provides a way for electrical detection of the orbital current in orbitronic device applications.

In the age of information technology, spin-based electronics has found wide application in data storage and processing. This field is facing increasing challenges that demand increasingly efficient materials for generation and manipulation of spin currents [1, 2]. It turned out that spin-orbit coupling (SOC) enables generation of nonequilibrium spin accumulation, e.g. by spin Hall effect (SHE) [3-5] and Rashba-Edelstein effect [6-9]. A reciprocal effect such as inverse SHE (ISHE) provides a way to detect the spin in an electrical way. Injection of the SOC-induced spin to a ferromagnet can exert torque on the magnetization, which is known as spin-orbit torque (SOT), an effect that originates from the exchange interaction between non-equilibrium spins and local magnetic moments [10-12]. Since SOC increases with the atomic number, heavy metals such as Pt, Ta and W have been widely used in spintronics investigation, making them prime candidates for devices used to manipulate the magnetization by electrical means only. In this scenario, light materials such as Cu are often overlooked, because they have negligible SOC.

However, recent studies have shown that naturally oxidized Cu films can significantly enhance SOT efficiency, which reveals a crucial role of an oxide interface [13-16]. This effect can be explained in the newly developed framework of orbital angular momentum (OAM) currents, which can be generated even in weak SOC materials. Apparently, OAM would not play any important role due to the angular momentum quenching in solids, however, it has been shown that the orbital Hall effect (OHE) can generate an OAM current in a transverse direction to an external electric field even if the OAM is completely quenched in the ground state [17-21]. At surfaces and interfaces, where the inversion symmetry is broken, orbital hybridizations induce a local electric dipole which influences the OAM polarization. Its interaction with structural asymmetry results in the orbital Rashba effect (ORE) [22-27], which induces a chiral OAM texture for electronic states in *k*-space. Thus, application of an external electric field can result in nonequilibrium OAM accumulation, which is known as the orbital Rashba-Edelstein effect (OREE) [27-28] – the orbital counterpart of the spin Rashba-Edelstein effect [6-8]. Since the orbital carries angular momentum, transfer of the OAM to the magnetization of a magnetic material provides an alternative mechanism to induce magnetization dynamics [29]. Such orbital torques (OTs) provide a promising route for magnetic nanodevices based on light elements.

Among different material candidates, a surface-oxidized Cu ($CuO_x$) is found to exhibit a strong OREE [27]. This is supported by recent papers on OT in heterostructures involving $CuO_x$ [27, 29, 30]. In particular, Ref. [29] demonstrated a novel route to achieve large torque efficiency using strong SOC of Pt in TIG/Pt/$CuO_x$ structure, where TIG = $Tm_3Fe_5O_{12}$. In Ref. [29], the OAM induced by the OREE is harnessed by the orbital-to-spin conversion in Pt which has strong SOC, and the resulting spin exerts torque on the magnetization of TIG. So far, we emphasize that there is still no experimental evidence of its reciprocal process, the inverse OT. We also note that Refs. [16,31] found negligible efficiency for the inverse OT.

In this Letter, we report an experimental observation of the inverse OT in heterostructures of YIG(40)/Pt($t_{Pt}$)/$CuO_x$(3) by means of spin pumping via ferromagnetic resonance (FMR) and longitudinal spin Seebeck effect (LSSE) in YIG, where the numbers in parentheses indicate the thickness of each layer in nanometers. When comparing with the result in YIG(40)/Pt($t_{Pt}$), we find substantial increase of the inverse SHE-like signal upon adding a $CuO_x$ capping layer, suggesting a crucial role of Pt/$CuO_x$ interface, where strong ORE exists. From this, we conclude that upon spin injection, nonequilibrium electronic states exhibit strong entanglement between the spin and orbital due to the strong SOC of Pt, and it is converted into charge current by the inverse OREE (IOREE) at Pt/$CuO_x$ interface. The efficiency of this process depends on two aspects: (1) the spin-orbital degree of entanglement that determines how efficiently the orbital current is conducted upward; (2) how strong is the IOREE, which is driven by the perpendicular E-field (due to the inversion symmetry breaking and/or any charge transfer process due to the oxidation) at the Pt/$CuO_x$ interface. Figure 1 presents a schematic illustration of the mechanism for converting spin-orbital entangled current into charge current by two mechanisms. The spin is injected from YIG and as the spin entangles with the OAM at Pt, it is converted into a charge current not only by the inverse SHE-OHE in Pt [Fig. 1(a)] but also via the IOREE at the $CuO_x$ interface [Figs. 1(b,c)]. We find that this contribution is much stronger than the "conventional" contribution by the ISHE in Pt. Our finding not only demonstrates the inverse OT unambiguously, but also sheds lights on the nonequilibrium spin-orbit physics. The YIG(40) film used in this work was grown by liquid phase epitaxy onto a 0.5 mm thick (111)-oriented $Gd_3Ga_5O_{12}$ (GGG). The Pt and Cu films were DC-sputter-deposited at room temperature in a working pressure of 2.8 mTorr and a base pressure of $2.0 \times 10^{-7}$ Torr or lower. The set of samples with the Cu capping layer was left outside the chamber for two days to naturally oxidize. All the YIG samples were cut from the same YIG(40) film in pieces with lateral dimensions of 1.5 x 3.0 $mm^2$ (see Ref. [32] for details).

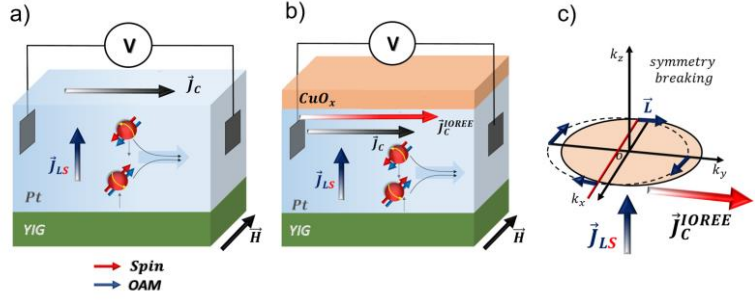

**Fig. 1.** Illustration of the inverse SHE-OHE process. a) Without CuO$_x$ layer, the only contribution is due to the conversion of $\boldsymbol{J}_{LS}$ in charge current. b) With CuO$_x$ layer, the additional $\boldsymbol{J}_C^{IOREE}$ appears driven by the interface Pt/CuO$_x$. c) IOREE in the presence of a chiral OAM texture in $\boldsymbol{k}$-space.

The investigation of the structural properties of the heterostructures was carried out by conventional X-ray diffraction (XRD), grazing incidence X-ray diffraction (GIXRD), and scanning electron microscopy (SEM) images. To investigate the oxidation process, we deposited a thick film of Cu(60) on top of the Pt(3) layer and left the sample exposed to air. The GIXRD pattern in Fig. 2(a) clearly shows the diffraction peaks characteristic of CuO$_x$ and of the polycrystalline Cu film with preferential texture oriented along the planes (111), (200), (220) and (202), as previously reported [33]. We deposited a capping layer of Pt(5) on the Cu surface and the GIXRD pattern is shown in Fig. 2(b) where no CuO$_x$ peaks were identified. Clearly the capping layer of Pt(5) prevents the Cu oxidation. The XRD spectrum also shows the peak corresponding to (111) Bragg reflections from the Cu film, demonstrating that the orientation of the GGG substrate strongly influences the orientation of the Cu grains. Figures 2(c,d) show SEM images of Cu/CuO$_x$ films deposited in GGG (111)/Pt(3) revealing a continuous film with different shapes of CuO$_x$ pyramids with dimensions smaller than 1 μm associated to the orientation of the GGG substrate.

We used FMR-driven spin-pumping (FMR-SP) technique excited at 9.41 GHz [34, 35] to investigate the interplay between orbital and spin currents in two series of heterostructures: YIG(40)/Pt($t_{Pt}$) and YIG(40)/Pt($t_{Pt}$)/CuO$_x$(3), where $0 \leq t_{Pt} \leq 7$ nm. For the second series of samples, a Cu(3nm) island with dimensions 1.5 x 2.0 mm$^2$ was deposited on top of the Pt layer. In this technique, illustrated in Fig. 3(a), a spin current is coherently injected through the YIG/Pt interface by the uniform precession of the YIG magnetization under FMR. The upward spin current density ($J_S^{SP}$), diffuses through the Pt layer generating a local electric field, $\boldsymbol{E}_{ISHE} \propto \boldsymbol{j}_S^{SP} \times \boldsymbol{\sigma}$, by means of the ISHE, where spin current polarization $\sigma \parallel \mathbf{M}_{YIG}$. By measuring the voltage difference ($V_{SP}$) produced between the two electrodes, we can define the SP signal as the current $I_{SP} = V_{SP}/R$, where $R$ is the electric resistance along the Pt layer. Figure 3(b) shows typical $I_{SP}$ signals for the sample YIG/Pt(2)CuO$_x$(3), which obey the equation $J_C^{SP} = (2e/\hbar)\theta_{SH}(J_S^{SP} \times \sigma)$ as expected, i.e., null at $\varphi = 90°$ (black), maximum positive at $\varphi = 0°$ (blue) and maximum negative $\varphi = 180°$ (red). Here $\varphi$ is the angle between the DC field and the direction of the voltage measurement. Inset of Fig. 3(b) shows the same behavior obtained for the sample YIG(40)/Pt(2). Figure 3(c) clearly shows the significant gain of the SP signal, for the sample YIG(40)/Pt(2)/CuO$_x$(3) (black) compared with the sample YIG(40)/Pt(2) (red).

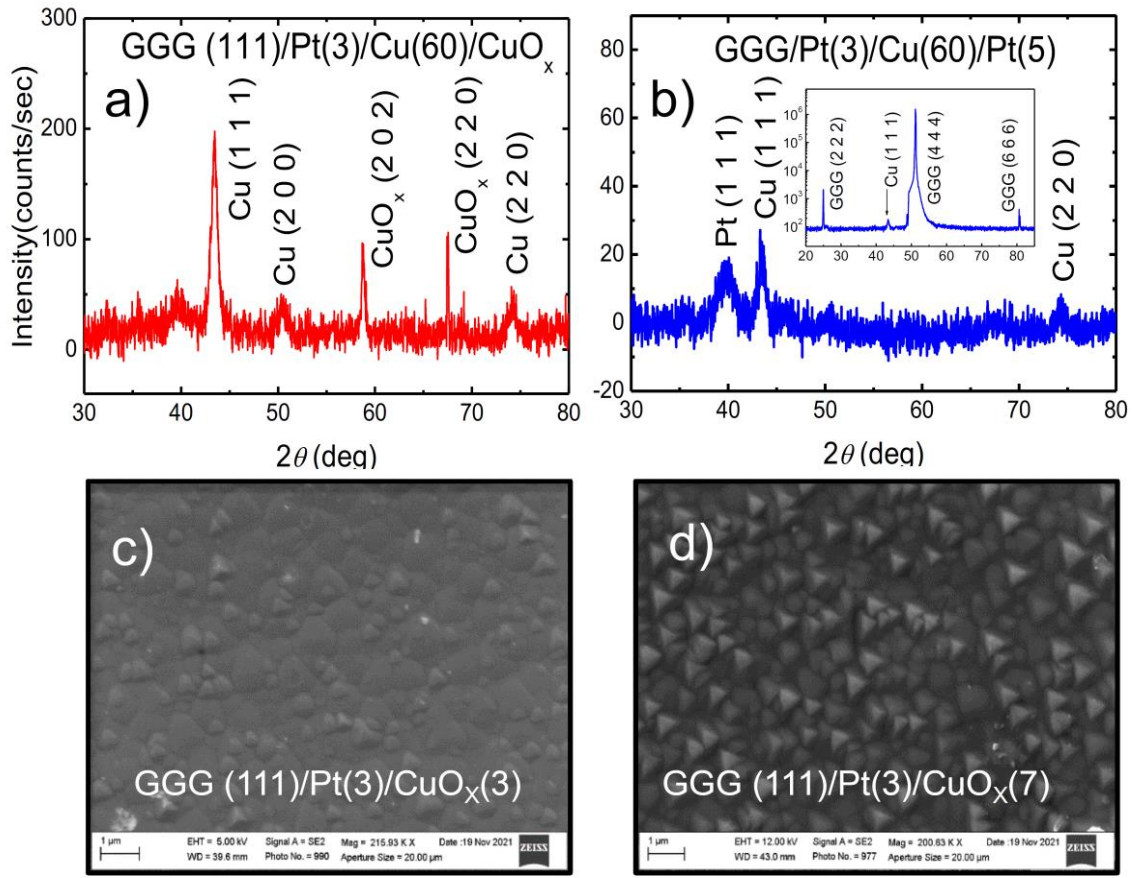

**Fig. 2**. a) Shows the GIXRD pattern of a film of Cu(60) that was left in contact with air for two days. The result shows the presence of strong CuO$_x$ peaks. In b), the GIXRD pattern clearly shows that a capping layer of Pt(5) prevents the Cu layer oxidation. Figs. c) and d) show SEM surface images of CuO$_x$(3) and CuO$_x$(7), respectively. The images reveal pyramidal structures associated to the orientation of the GGG substrate. Inset of b) shows the conventional XRD pattern of GGG/Pt(3)/Cu(60)/CuO$_x$, exhibiting reflections associated with the (222), (444) and (666) crystal planes of GGG as well as (111) texture of the Cu layer.

While the bare YIG presented an FMR linewidth of $\Delta H = 2.63$ Oe, it increases to 2.84 Oe after deposition of Pt and increases for 3.06 Oe after deposition of $CuO_x$ on top of Pt, which characterizes the transfer of spin angular moment from YIG to the Pt layer (inset of Fig. 3(c)). To confirm the effect of the Cu oxidation, we prepared a sample in which the Cu layer is protected by a layer of MgO(5). The results of Fig. 3(d), shows that the $I_{SP}$ signals for the samples YIG(40)/Pt(2)/Cu(3)/MgO(5) (blue) and YIG(40)/Pt(2) (red), are equivalent, thus confirming that the enhancement shown in Fig. 3(c) comes from the Pt(2)/$CuO_x$(3) interface. Figure 3(e) shows the dependence of the peak value of $I_{SP}$, as a function of the rf power used to excite the FMR, for both samples [32]. By comparing the slopes of the black and red lines, we observed that the gain in $I_{SP}$, due to the presence of $CuO_x$(3), is around 5 times. The dependence of $I_{SP}$ as a function of the Pt layer thickness, with a rf power of 110mW, for both set of samples is shown in Fig. 3(f). The $I_{SP}$ signal for the sample YIG(40)/Pt($t_{Pt}$) (red) exhibits the usual behavior, i.e., increases as the $t_{Pt}$ increases reaching the saturation for $t_{Pt} > 4$ nm. On the other hand, the dependence of $I_{SP}$ on $t_{Pt}$ for the set of samples with a capping layer of $CuO_x$(3) (black), increases sharply reaching a maximum at $t_{Pt} \sim 3$ nm, then decreases to the same range of saturation values observed for the YIG(40)/Pt($t_{Pt}$) samples. The solid lines in Fig. 3(f) are guide to the eyes. The unusual dependence of $I_{SP}$ on $t_{Pt}$, for the samples YIG(40)/Pt($t_{Pt}$)/$CuO_x$(3), is explained as due to the interplay between the spin and orbital states that is mediated by the strong SOC of Pt. The spin-orbital entangled current ($\boldsymbol{J}_{LS}$) is partially converted into charge current by means of the inverse spin and orbital Hall effects of Pt and by the strong IOREE at the Pt/$CuO_x$ interface. As the Pt layer thickness increases, the $\boldsymbol{J}_{LS}$ current no longer reaches the Pt/$CuO_x$ interface due to the finite diffusion length of the spin-orbital entangled current in Pt, thus causing the measured voltage to decrease.

To confirm the results obtained by SP, we performed spin injection measurements using the LSSE technique on the same two sets of samples. In LSSE, the application of a temperature gradient in a magnetic material generates a spin current along the direction of the temperature gradient, which is the magnetic analog of the thermoelectric Seebeck effect [36]. Figure 4(a) shows schematically the LSSE setup, in which a thermal gradient is applied perpendicular to the sample plane [32]. It highlights the perpendicular temperature gradient $\boldsymbol{\nabla}T$ and the upward spin-orbital entangled current ($\boldsymbol{J}_{LS}$), the ISHE-like charge currents in Pt ($J_C$) due to the inverse spin and orbital Hall effects, as well as the OREE current on the Pt/CuO$_x$ interface.

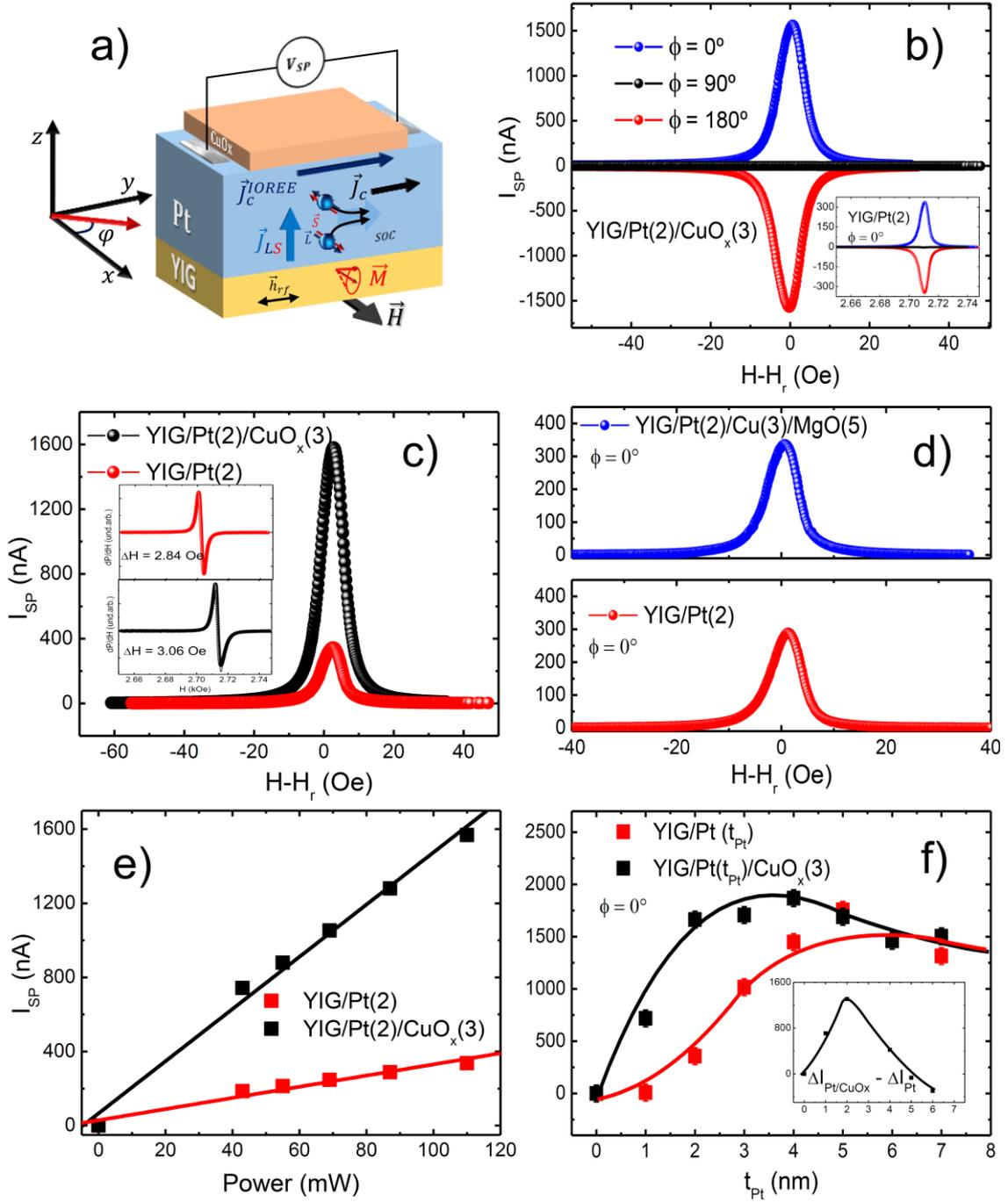

**Fig. 3** - a) Schematically shows the SP configuration. b) Shows the typical $I_{SP}$ signals for the samples with and without the CuO$_x$ cover layer (inset). c) Shows the comparison of the SP signals of the samples with (black) and without (red) CuO$_x$ capping layer, respectively, obtained for $P_{rf} = 110$mW. The inset show the derivative of the FMR absorption signal for both samples. d) Shows $I_{SP}$ signals for the samples YIG(40)/Pt(2)/Cu(3)/MgO(5) (blue) and YIG/Pt(2) (red), which confirm that the enhancement occurs only when the Cu cover layer is oxidized. e) Shows the dependence of $I_{SP}$ signals, as a function of the rf power, for the samples YIG/Pt(2) (red) and YIG/Pt(2)/CuO$_x$(3) (black). f) Shows the dependence of $I_{SP}$ as a function of $t_{Pt}$, for the samples YIG/Pt(2) (red) and YIG/Pt(2)/CuO$_x$(3) (black). The solid lines are guide to the eyes. The inset shows the difference between the data of the samples with and without the CuO$_x$ cover layer, which reaches a maximum for $t_{Pt} \approx 2nm$.

Figure 4(b) shows LSSE signals, where thermal voltages ($V_{\text{LSSE}}$) are measured along sample plane for heterostructures of YIG(40)/Pt(2) (red) and YIG(40)/Pt(2)/CuO$_x$(3) (black). From negative to positive field sweep, the perpendicular temperature gradient is fixed at the same value for the measurements of both samples. Surprisingly, the measured LSSE signal value for the sample with a CuO$_x$ capping layer increased more than two times compared to the sample without CuO$_x$. Figure 4(c) shows the LSSE signals of YIG(40)/Pt(2)/CuO$_x$(3) for various temperature differences ($0K \leq \Delta T \leq 20K$) between the heat baths placed on the lower and upper surfaces of the sample [32]. The LSSE signal amplitude ($\Delta I_{\text{LSSE}}$), defined in Fig. 4(b), increases linearly as a function of $\Delta T$, as summarized in Fig. 4(d) for the heterostructures of YIG(40)/Pt(2)/CuO$_x$(3) (black) and YIG(40)/Pt(2) (red), respectively. The linear fits of Fig. 4(d) were obtained by means of $\Delta I_{\text{LSSE}} = -\alpha_{\text{FM}} \Delta T \cos\varphi$, where $\alpha_{\text{FM}} = \frac{S_{\text{FM}}}{R}\left(\frac{w_{\text{Pt}}}{t_{\text{YIG}}+t_{\text{GGG}}}\right)$ [37, 38]. Here, $S_{\text{FM}}$ is the spin-Seebeck coefficient, $w_{\text{Pt}}$ is the distance between the electrical contacts, $\varphi$ is the azimuthal angle as defined in Fig. 4(a), and $t_{\text{YIG}}$ and $t_{\text{GGG}}$ are the thicknesses of the YIG and GGG layers, respectively [32]. From the linear fits shown in Fig. 4(d) we obtained $\alpha_{\text{FM}}^{(1)}/\alpha_{\text{FM}}^{(2)} \cong 2.6$, where $\alpha_{\text{FM}}^{(1)}$ and $\alpha_{\text{FM}}^{(2)}$ are the linear coefficients for the black and red lines, respectively. Thus, the presence of CuO$_x$ increased the LSSE signal by a factor of 2.6. Figures 4(e,f) show the dependence of $\Delta I_{\text{LSSE}}$ for both series of samples as a function of $t_{\text{Pt}}$ for $\Delta T = 27$ K and $\Delta T = 3$ K, respectively. As expected, the dependence of $\Delta I_{\text{LSSE}} \times t_{\text{Pt}}$ for YIG(40)/Pt($t_{\text{Pt}}$) is given by $\Delta I_{\text{LSSE}} = \beta \lambda_N \tanh(t_{\text{Pt}}/2\lambda_N)$, where $\beta$ depends on the interface YIG/Pt properties and $\lambda_N$ is the spin diffusion length in the Pt layer [32,37]. $\Delta I_{\text{LSSE}}$ increases monotonically, reaching the saturation for $t_{\text{Pt}} > 4$ nm, and the red line corresponds to the best fit to the data with $\lambda_N = 1.6 \pm 0.2$ nm, which is in the range of reported spin diffusion lengths for Pt that span an order of magnitude, ranging from just over 1 nm to 10 nm [35, 39]. On the other hand, the results for the samples YIG(40)/Pt($t_{\text{Pt}}$)/CuO$_x$(3), indicated in Figs. 4(e,f) by the black symbols, shows the same behavior of the SP signal of Fig. 3(f). Black lines are guide to the eyes. The LSSE signal rapidly increases as a function of $t_{\text{Pt}}$, reaches a maximum for $t_{\text{Pt}} \approx 3$ nm and then decreases to values of the same order as the saturation values obtained for the YIG(40)/Pt($t_{\text{Pt}}$) samples.

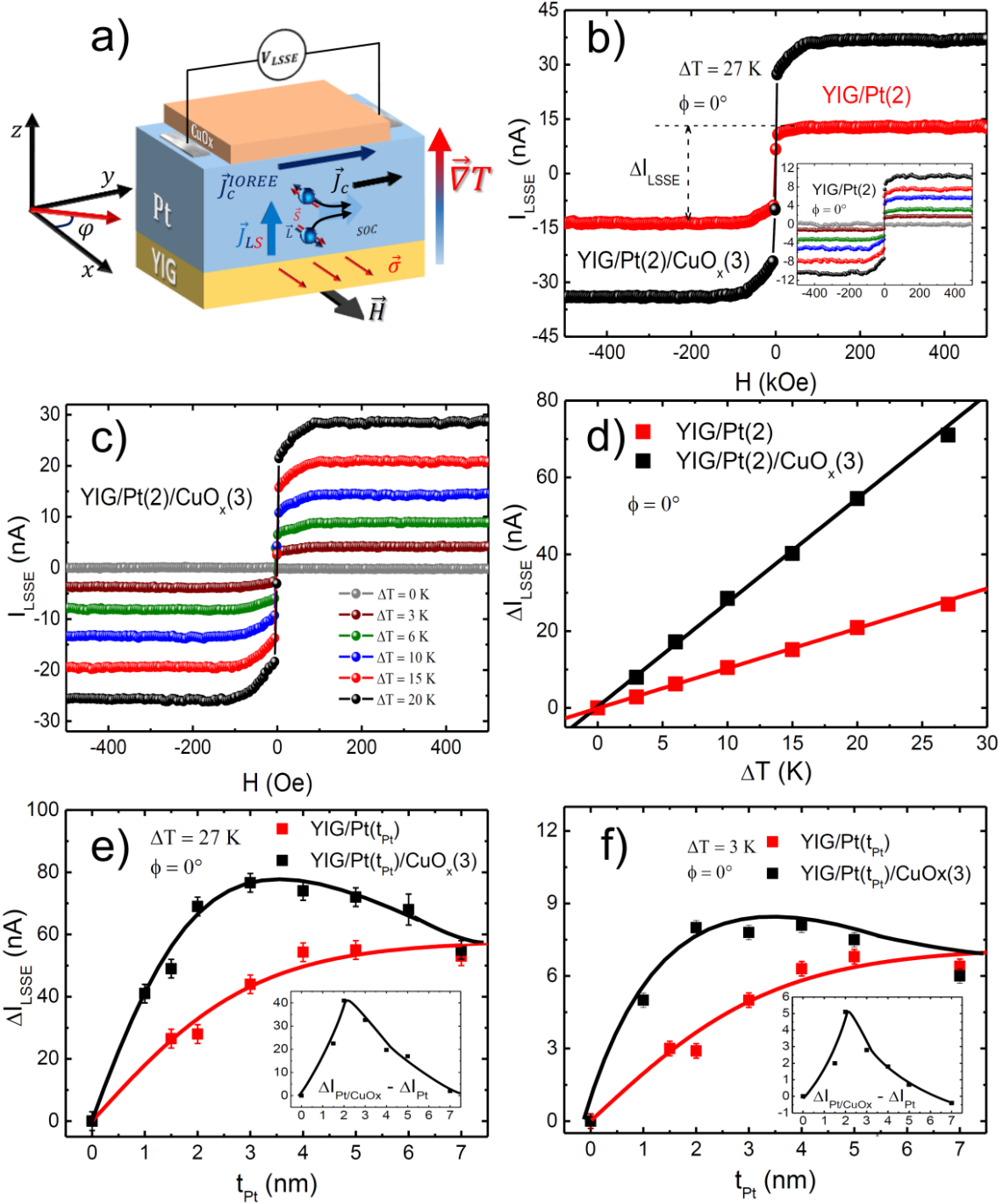

**Fig. 4** - a) Schematically shows the LSSE configuration. b) Shows the field sweep of the LSSE signal for the samples with (black) and without (red) a capping layer of $CuO_x(3)$, for $\Delta T = 27K$. With the deposition of the $CuO_x$ layer, the LSSE signal obtained a gain of order 2.6. c) Shows the field scans curves for values of $\Delta T$ ranging from 0K to 20K, for the sample with the cover layer of $CuO_x$. The inset of 4 b) shows the same curves obtained for the sample YIG(40)/Pt(2). d) Shows the dependence of $\Delta I_{LSSE}$ as a function of $\Delta T$ for the samples YIG/Pt(2)/$CuO_x$(3) (black) and YIG/Pt(2) (red), where $\Delta I_{LSSE}$ is defined in b). Figs. 4 e) and f) show the dependence of $\Delta I_{LSSE}$ as a function of $t_{Pt}$ for $\Delta T = 27K$ and $\Delta T = 3K$, respectively. The red symbols are the data for the sample YIG/Pt($t_{Pt}$) and the black symbols are the data for the sample with the capping layer of $CuO_x$. The insets of Figs. 4 e) and f) shows the difference between the data of the samples with and without the $CuO_x$ cover layer, which reaches a maximum for $t_{Pt} \approx 2nm$.

This very peculiar behavior can also be explained as due to the interplay between the electron spin and orbital degrees of freedom. In the LSSE, an upward spin current, $J_S^{\mathrm{LSSE}} \parallel \nabla_z T$, driven by the $\nabla_z T$, is injected through the YIG/Pt interface. Due to the strong SOC of Pt, the out-of-equilibrium spin states, thermally pumped into Pt, give rise to a spin-orbital entangled current ($J_{\mathrm{LS}}$). As the entangled spin-orbital current propagates upward through the Pt layer, it is converted into charge current, either by the inverse SHE or inverse OHE, as well as by the strong OREE at the Pt/CuO$_x$ interface. It is important to mention that preliminary results performed in YIG/Pt(4)/AlO$_x$(3) also exhibited a gain of more than twice compared to YIG/Pt(4) (See Fig. S4 in the Supplemental Material).

The interaction between charge, spin, and orbital degrees of freedom, triggered by the FMR-driven or thermal-driven spin pumping, represents a unifying principle to explain the experimental results reported in this work. By injecting the pure spin current from the FMR-SP or LSSE from YIG into Pt, the nonequilibrium electronic states have strong entanglement between the spin and orbital and carries spin-orbital entangled current. In Pt, a part of the spin-orbital entangled current is converted into charge current. This occurs either by the inverse SHE or inverse OHE. We note that this has been conventionally interpreted solely in terms of the inverse SHE, neglecting the orbital contribution. Meanwhile, as the spin-orbital entangled current propagates across the Pt layer and reaches the CuO$_x$ interface, it is converted to charge current via the inverse OREE. However, if the Pt layer is thicker than the relaxation length for the spin-orbital entangled current, it cannot reach the CuO$_x$ interface and only the inverse SHE/OHE contribution contributes. This explains why the efficiencies for YIG/Pt and YIG/Pt/CuO$_x$ converge to the same value for $t_{\mathrm{Pt}} > 7\mathrm{nm}$.

In conclusion, we used LSSE and FMF-SP techniques to investigate the interplay between spin, orbital and charge degrees of freedom in heterostructures of YIG(40)/Pt($t_{\mathrm{Pt}}$)/CuO$_x$(3). Due to the strong SOC of Pt, the spin states, pumped through the YIG/Pt interface, entangle with the local orbital states, thus generating an upward pure spin-orbital current ($J_{LS}$) without flow of charge. Part of this current is converted, within the Pt, into a transverse charge current by either the inverse-SHE effect or the inverse-OHE effect. Part of the remaining spin-orbital current that flows upward, is transformed into a transverse charge current by means of the inverse OREE. This current is added to the previous charge current, thus increasing the resulting charge current. The dependence of the charge current as a function of Pt layer thickness, provides a clear picture of the phenomenon of converting spin-orbital current to charge current. As the Pt layer thickness becomes larger than the spin-orbital diffusion length, the $J_{LS}$ no longer reaches the Pt/CuO$_x$ interface, and the IOREE mechanism ceases to occur. Therefore, the charge current value is reduced to the saturation values of the charge currents generated only by the inverse SHE combined with the inverse OHE. Certainly, the results reported here open new avenues to understand the basic mechanisms underlying the spin-orbital entanglement phenomena.

**Acknowledgements**

This research was supported by Conselho Nacional de Desenvolvimento Científico e Tecnológico (CNPq), Coordenação de Aperfeiçoamento de Pessoal de Nível Superior (CAPES), Financiadora de Estudos e Projetos (FINEP), Fundação de Amparo à Ciência e Tecnologia do Estado de Pernambuco (FACEPE), Universidade Federal de Pernambuco, Fundação de Amparo à Pesquisa do Estado de Minas Gerais (FAPEMIG) - Rede de Pesquisa em Materiais 2D and Rede de Nanomagnetismo.